\documentclass[pdflatex]{sn-jnl}

\newcommand{\half}{\frac{1}{2}}
\newcommand{\an}{\hat{a}}
\newcommand{\ad}{\hat{a}^\dagger}
\newcommand{\phicap}{\hat{\phi}}
\newcommand{\ncap}{\hat{n}}
\newcommand{\Hcap}{\hat{H}}

\usepackage{graphicx}
\usepackage{braket}
\usepackage{physics}
\usepackage{amsmath}
\usepackage{mathtools}
\usepackage{braket}

\jyear{2021}%

\begin{document}

\title[Quantifying the effects of dissipation and temperature ..... ]
      {Quantifying the effects of dissipation and temperature on dynamics of a
        superconducting qubit-cavity system}

\author*[1,2]{\fnm{Prashant} \sur{Shukla}}\email{pshukla@barc.gov.in}

\affil[1]{\orgdiv{Nuclear Physics Division}, \orgname{Bhabha Atomic Research Centre},
  \orgaddress{\city{Mumbai}, \postcode{400085}, \country{India}}}

\affil[2]{\orgname{Homi Bhabha National Institute},
  \orgaddress{\street{Anushakti Nagar}, \city{Mumbai}, \postcode{400094}, \country{India}}}

\abstract
{The superconducting circuits involving Josephson junction offer
macroscopic quantum two-level system (qubit) which are coupled to cavity
resonators and are operated via microwave signals. 
 In this work, we study the dynamics of superconducting qubits coupled to
a cavity with including dissipation in a subkelvin temperature domain.
In the first step, a classical Finite Element Method is used to simulate
the cavities and basic circuit elements to model Josephson junctions.
 Then the quantization of the circuit is done to obtain the full Hamiltonian of
 the system using energy partition ratios of the junctions.
 Once the parameters of Hamiltonian are obtained,
 the dynamics is studied via Lindblad equation for an open quantum system
 using a realistic set of dissipative parameters and include temperature effects.
 Finally, we get frequency spectra and/or dynamics of the system with time which
 have quantum imprints. 
 Such devices work at tens of milli Kelvins and we search for a set of
 parameters which could enable to observe quantum behaviour at temperatures
 as high as 1 K. 
}


\maketitle

\section{Introduction}

 Quantum technologies exploit the properties of quantum phenomena, such as
 superposition and entanglement \cite{Dowling_2003,nielsen_chuang_2010}.
 Quantum computers,
 quantum communications and quantum sensors are some of the fast-developing
 areas involving quantum technologies \cite{Ladd2010,Degen2017}.
A classical bit can take two states; 0 and 1 but a quantum bit (qubit) can be described
as a superposition of two basis states $\ket{0}$ and $\ket{1}$. 
While $N$ classical bits represent one of the $2^N$ possible states,
$N$ quantum bits can represent all of the $2^N$ possible states and one can operate
on all states simultaneously.
 The basic building block of a quantum computer is a macroscopic
quantum two-level system (qubit).
To have a quantum computer, one should be able to
create and manipulate the quantum states, measure the quantum state and
build a multiqubit entangled system. 
 A qubit can be formed using superconducting
 circuits involving Josephson junction
 \cite{Nakamura1999,Koch2007,Girvin_2009,Blais_2007, martinis2020, Krantz_2019}.
 The qubits are coupled to cavity
resonators which could be 3-dimensional or planar \cite{Naghiloo_2019}.
The quantum states are manipulated using microwave signals via cavity resonator. 

The design of a superconducting quantum device involves three steps, first,
a classical Finite Element Method is used to
simulate the cavities and basic circuit elements to model Josephson junctions.
Then the quantization of the circuit is done to obtain the full Hamiltonian of
the system using energy partition ratios (EPR) of the junctions \cite{Minev_2019}.
In the third step, a quantum mechanical equation such as Lindblad equation 
\cite{Manzano_2020} is solved to obtain the energy levels and
probabilities of qubit states.
 We consider 3-dimensional cavities both rectangular and cylindrical which
are coupled to qubit. The parameters of Hamiltonian of the system
such as qubit cavity coupling are obtained using EPR method~\cite{Minev_2019}. 
 An open quantum system formalism is implemented to study the effect of dissipation
and finite temperature on frequency and time spectra.

In this work, we give all ingredients required to calculate the dynamics of qubit
coupled with a cavity. 
Section 2 gives basic formalism for superconducting LC circuit, section 3 describes
Josephson junction and construction of transmon. Two types of resonators
are discussed in section 4. Section 5 describes how the parameters of Hamiltonian for a qubit-cavity
system can be obtained using EPR method. Section 6 describes Lindblad equation,
entanglement and treatment of dissipation and thermal effect. Section 7 gives
the effect of bath temperature on vacuum Rabi oscillations. Section 8 describes how a qubit is
driven using microwave signals. Section 9 gives the measurement of frequency shift. Summary
is given in section 10.

\section{Superconducting circuits}
Quantum mechanics is usually invoked when dealing with atomic or microscopic world.
Superconducting circuits can offer macroscopic quantum systems, the parameters of which
are not God-given constants but can be tailored by the design of the system.
The most basic component is a superconducting LC circuit which  works as a
quantum harmonic oscillator.
 The Hamiltonian of the LC circuit is given by \cite{Krantz_2019}
\begin{eqnarray}
  H & = \half C\,V^2 + \half L\, I^2  \nonumber \\
   & = {1 \over 2C } Q^2 + {1 \over 2L} \Phi^2,
\end{eqnarray}
where $Q$ and $\Phi$ are charge and flux, respectively which can be converted to dimension less
quantities using charge quantum $e$ and flux quantum $\Phi_0=\hbar/(2e)$ as
\begin{eqnarray}
n=\frac{Q}{2e}, \,\,\, \phi=\frac{\Phi}{\Phi_0}.
\end{eqnarray}
In terms of $n$ and $\phi$ the Hamiltonian can be written as
\begin{eqnarray}
 H & =  4 E_C \, n^2 +  \half E_L  \, \phi^2.
\end{eqnarray}
Here, $E_C = e^2/(2C)$ is charging energy per electron 
and $E_L = \Phi_0^2/L$ is inductive energy.
The Hamiltonian operator can be obtained in terms of Ladder operators as
\cite{nielsen_chuang_2010}
\begin{eqnarray}
  \Hcap & = & 4 E_C \, \ncap^2 + \half E_L \phicap^2 \\
  & = & \hbar \sqrt{8E_CE_L} \left( \ad \an + \half \right).
\end{eqnarray} 
The ladder operators are defined by 
\begin{eqnarray}
  \phicap &= \sqrt{\xi } \, (\an + \ad), \,\,\,\,\,
  \ncap &= \frac{i}{2\sqrt{\xi}} \, (\an - \ad), \,\,\,\,\,
  \xi  \,\, =  \sqrt{2E_C/E_L}.
\end{eqnarray}
The energy levels of harmonic oscillator are obtained by solving the Schrodinger
equation and are given by 
$ E_n =  \hbar \omega_r \left(n + \half \right), \,\,\,\,\, n=0, 1, 2, ..... $,
$\omega_r/2\pi = \sqrt{8E_LE_C}/h$.

\section{Transmon}

A Josephson junction is the basic element used in superconducting qubits.
The inductance of the Josephson junction is variable and it is
shunted by a large capacitance $C_s$ to make transmon with the total
capacitance given by $C=C_j+C_s$ where $C_j$ is the junction capacitance\cite{Koch2007}.
The Josephson relations are given by \cite{Krantz_2019}
\begin{eqnarray}
  I &=& I_C\, \sin \phi, \,\,\,\,\,\, ( {\rm linear \,\, case} \,\,\ I = \Phi/L) \nonumber \\
  V &=& \Phi_0 \,\dot{\phi}.
\end{eqnarray}
Josephson inductance is given by 
\begin{eqnarray}
  L = \frac{V}{\dot{I}} = \frac{\Phi_0}{I_C} \frac{1}{\cos\phi}.
\end{eqnarray}
Here,  $I_C$ is the critical current. 
The energy stored in the junction (Josephson energy) is
\begin{eqnarray}
  E(t) = - E_J \cos\phi .
\end{eqnarray}
Here $E_J  =  I_C \Phi_0$ and $E_C = e^2/2C$.
The transmon Hamiltonian can be written as 
\begin{eqnarray}
  H = 4 E_C n^2 - E_J \cos\phi. 
\end{eqnarray}
To have an idea of typical values, for $C = 0.1$ pF,
$I_C=55$ nA, we get $E_C/2\pi$ = 223 MHz and $E_J/2\pi$ = 27.31 GHz.
The Transmon Hamiltonian can be expanded as \cite{Didier_2018}

\begin{eqnarray}
  H & = & 4 E_C \, n^2 + \half E_J \phicap^2 - \frac{1}{4!} E_J \phicap^4,
        + \frac{1}{6!} E_J \phicap^6 - ....... ,
\end{eqnarray}
where
\begin{eqnarray}
  \phicap = &\sqrt\xi \, (\an + \ad),  \,\,\,\,\,\,  \xi =  \sqrt{\frac{2E_C}{E_J}}.
\end{eqnarray}
It requires operator expansion for each term in the power of $\phicap$
for $E_J>>E_C$ (Transmon limit).
The energy levels of transmon are not equidistant. The difference between
the first two energy levels is $\omega_q=E_1-E_0$ and the anharmonicity is
defined as $\alpha=(E_1-E_0)-(E_2-E_1)$. The frequency of the driving
microwave signal should be equal to $\omega_q$ and the width should be less than
$\alpha$.
Table~\ref{table_transmon} gives transmon parameters, $\omega_q=2\pi \,\sqrt{8\,E_C E_J}$
and $\alpha$ for $E_J=27.31$ GHz and $E_J/E_C=122.47$ for 3 orders of perturbation theory.

\begin{table}
\centering
  \caption{Transmon parameters, $\omega_q=2\pi \,\sqrt{8\,E_C E_J}$ and $\alpha$ for
   $E_J=27.31$ GHz and $E_J/E_C=122.47$. } 
\begin{tabular}{ | c | c | c |}
\hline 
Order & $\omega_q/2\pi$ (GHz)  & $\alpha/2\pi$ (MHz)  \\
\hline 
0  &   6.982   &   0.0  \\
1  &   6.759   &   223.1  \\
2  &   6.752   &   239.1  \\
\hline
\end{tabular}
\label{table_transmon}
\end{table}

For a more elaborate design numerical simulations are performed.
One can obtain zero point fluctuations $\sqrt{\xi}$ from Energy Participation Ratio (EPR)
method as \cite{Minev_2019}, 
\begin{eqnarray}
  \xi_{mJ} = p_{mJ} \,{\frac{\hbar\omega_m}{2E_J}},
\end{eqnarray}
where $p_{mJ}$ is the Energy Participation Ratio defined as the ratio of inductive
energy stored in junction $J$ to the inductive energy stored in the mode $m$.

 The critical current in the Josephson junction is given by \cite{ambegaokar1963}
\begin{eqnarray*}
 I_C = \frac{\pi \Delta(0)}{2eR_n}.
\end{eqnarray*}
Here, $\Delta(0)$ is the superconducting gap at zero temperature and is
$\sim 170 \, \mu$eV for Aluminum. $R_n$ is the resistance of the oxide layer
which depends on the junction area and thickness of the layer.
Its value is obtained as $4.87 \, k\Omega$ using $I_C=55$ nA for 
Al$_2$O$_3$ layer with resistivity $\rho_{Al2O3} = 10^{11} \Omega\, m$.
Making a Josephson junction requires a capability of few tens of nanometer pattern size
and variable angle electron beam evaporation is needed.
For a capacitor with pads 400 $\mu$m $\times$ 600 $\mu$m with a gap of 200 $\mu$m,
the capacitance is calculated as $C = 0.0866$ pF.

\begin{figure}
\centering
  \includegraphics[width=0.49\textwidth]{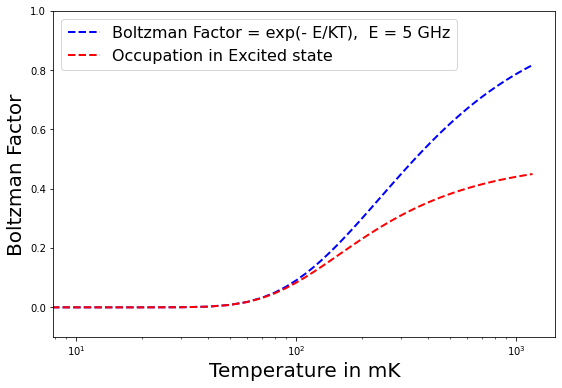}
  \includegraphics[width=0.49\textwidth]{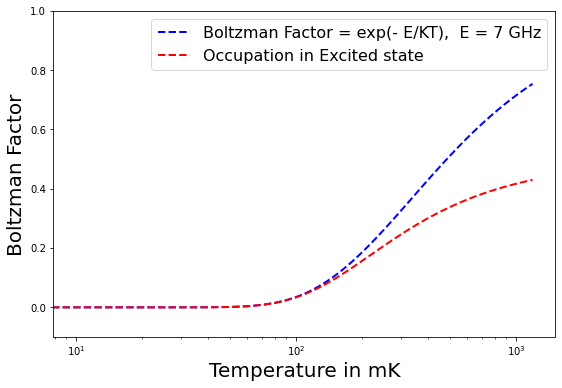}
\caption{The Boltzmann factor and occupation probability of excited state as a function of
temperature for two values of difference of energy levels between two states.}
\label{fig_boltz}
\end{figure}

The superconducting qubits are operated in milli Kelvin range since the energy
gap ($\sim \mu$eV) between qubit levels has to be smaller than the thermal excitation.
The probability of gaining or losing a photon from a thermal bath is given by
\begin{eqnarray}
  F =   e^{-\hbar\omega/kT}, \,\,\,\,\,\,\,\,
  P_{ex} = \frac{F}{1+F}. 
\end{eqnarray}
Figure~\ref{fig_boltz} shows the 
Boltzmann factor $F$ and occupation probability $P_{ex}$ of excited state as a function of
temperature for two values of difference of energy levels between the two states of a qubit.
These correspond to typical frequencies involving superconducting qubits. 
The quantum devices operate at tens of millikelvins which can go up to 100-200
millikelvins with small noise.
 The main components of a quantum device are qubits coupled to a resonator 
which is driven by microwaves.

\section{The resonator}

 The first step of design is a classical simulation of the resonator and transmon model.
Many kinds of resonators can be considered \cite{Reagor_2016}. We consider a rectangular resonator
with dimensions 36 mm $\times$ 6 mm $\times$  22 mm.  
Aluminium EC grade (99.7\%) with surface roughness, 0.5 microns is used in the simulation. 
The frequency is given by 
\begin{eqnarray}
f = {\omega_r \over 2 \pi} \, = \, {c\over 2} \, \sqrt{  \left({n_x \over L_x}\right)^2 +  \left({n_y \over L_y}\right)^2 + \left({n_z \over L_z}\right)^2  }.
\end{eqnarray}
TE$_{110}$ mode is 8 GHz and the finite element (FE) simulation gives 7.75 GHz with
a quality factor of 4268.
For Al 6061 (97.9\% purity) the quality factor is reduced.  
A transmon can be modeled as a lumped element with large capacitive pads on a silicon
substrate. The junction is modeled as a rectangular sheet (200 $\mu$m $\times$ 50 $\mu$m)
with a polyline with the current flow direction. An inductance is assigned to the junction.


We also use a $\lambda/4$ type of resonator. Height of stub is taken as 8 mm with
inner radius 1 mm. Height of cylinder is taken as 30 mm with a radius 7 mm. 
  One can obtain a rough estimate by $f = {c/4L}$.
  The FE simulations give 7.6 GHz with $Q$ factor as 2909. 


Dissipation ($\kappa$) in resonators is 
inversely related to Quality Factor $Q = \omega/\kappa$.
Dissipation for a rectangular cavity depends on three factors;

\begin{itemize} \item Dielectric loss which is due to oxide layer with thickness $t$.

\item Conductor loss which depends on surface penetration depth ($\lambda$). 

\item Seam loss which is due to seam location. The seam loss is zero if the seam is
  exactly at the middle. It depends on machining tolerance $\delta x_0$. 
\end{itemize}

Dissipation for $\lambda/4$ cavity depends on
\begin{itemize}
\item  Conductor loss: $1/Q_C= \frac{2 \sqrt{\pi\,f\mu\sigma} \,\ln(b/a)} { {1\over a} + {1\over b}}$, \\
  $a$ is the radius of the inner cylinder and $b$ is the radius of the outer cylinder,
  $\mu$ is the permeability and $\sigma$ is the conductivity of the conductor. 
\item  Dielectric loss: $1/Q_d = 1/\tan(\delta)$.
\end{itemize}

\section{Quantum simulation}
 After the classical simulation of resonator and transmon,
the electromagnetic circuit is then quantized and Hamiltonian parameters are
obtained using Energy Partition Ratio method.

Hamiltonian for a single qubit-cavity system is given by \cite{Jaynes_1963,Shore_1993}
\begin{eqnarray}
H &=& H_r + H_q + H_{int} \nonumber\\
 &=& \omega_r \ad\an - \half \omega_q \sigma_{z} + g\, (\an\sigma_+ + \ad\sigma_-).
\end{eqnarray}
where $\omega_q$ is qubit frequency and $\omega_r$ is cavity frequency and $g$ is
the coupling strength between them. 
This Hamiltonian can be solved in two modes. The polariton mode where 
$\triangle = \omega_q - \omega_r \simeq 0$ and dispersive mode, where 
$\chi = g^2/\triangle << 1$.

Hamiltonian for Qubit-Cavity system in dispersive limit is given by
\cite{schuster2007,Boissonneault_2009}
\begin{eqnarray}
 H & = & \omega_r \ad\an - \half (\omega_q-\alpha/2)\sigma_z + \chi(\ad\an+1/2) \sigma_z.
\end{eqnarray}
Here $\alpha$ is self kerr and $\chi=g^2/\triangle$ is cross kerr. 

 The Hamiltonian for two coupled qubits is given by 
\begin{eqnarray}
H = - \half \omega_1 \sigma_{z1} - \half \omega_2 \sigma_{z2} + g\, \sigma_{y1} \sigma_{y2}. 
\end{eqnarray}
Here $\omega_1$ and $\omega_2$ are the frequencies of the two transmons and
$g$ is the coupling strength between them. 
The parameters of the Hamiltonian are obtained by pyEPR simulation \cite{Minev_2019}.
The zero point fluctuations $\sqrt{\xi_{mJ}}$ are related to the $p_{mJ}$ which is the
 Energy Participation Ratio defined as \cite{Minev_2019}

\begin{equation}
  p_{mJ} = \frac{{\rm Inductive \,\, energy \,\, stored \,\, in \,\, junction \,\,} J}
  {{\rm Inductive \,\, energy \,\, stored \,\, in \,\, the \,\, mode \,\,} m}.
\end{equation} \\

\begin{figure}
\centering
  \includegraphics[width=0.45\textwidth]{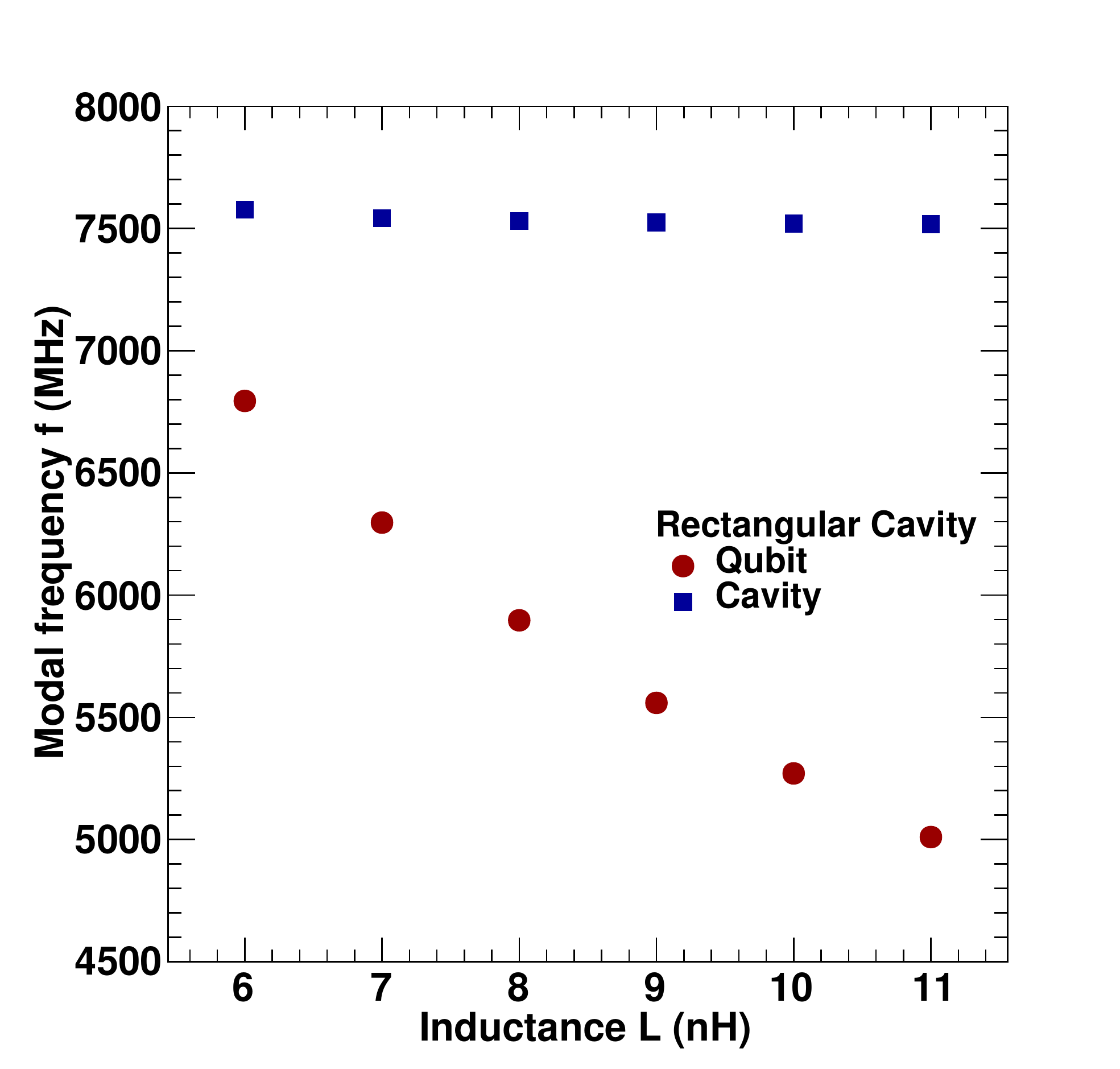}
  \includegraphics[width=0.45\textwidth]{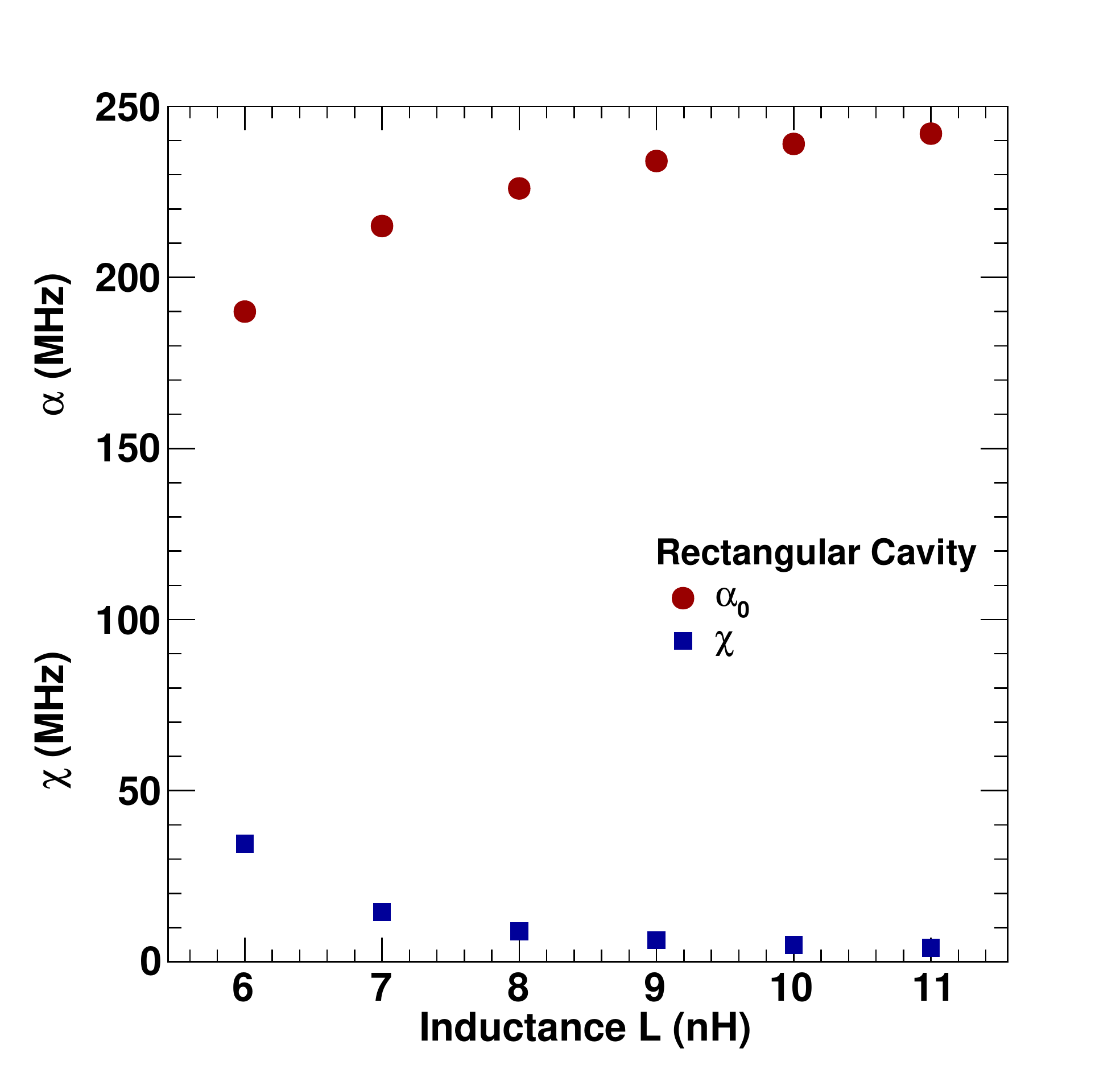}
  \caption{pyEPR simulations of a qubit coupled to a rectangular cavity showing modal 
frequencies (MHz), Anharmonicity $\alpha$ (MHz) and cross-kerr frequency $\chi$ (MHz) as a function of
different values of $L$.
}
\label{fig_kerrrect1}
\end{figure}

Figure~\ref{fig_kerrrect1} shows pyEPR simulations of a qubit coupled to a rectangular
cavity giving modal 
frequencies (MHz), Anharmonicity $\alpha$ (MHz) and cross-kerr frequency $\chi$ (MHz) as a function of
different values of $L$.
 For input parameter $L_J=6$ nH, $E_J=27.31$ GHz, the EPR value is 0.8 and 
$\omega_r/2\pi=$ 7.577 GHz, $\omega_q/2\pi=$  6.794 GHz, $\alpha/2\pi=$ 190 MHz,
 $\chi/2\pi=$ 34.4 MHz which gives $g/2\pi = \sqrt{\chi\triangle}=$ 181 GHz.
 
\begin{figure}
\centering
  \includegraphics[width=0.45\textwidth]{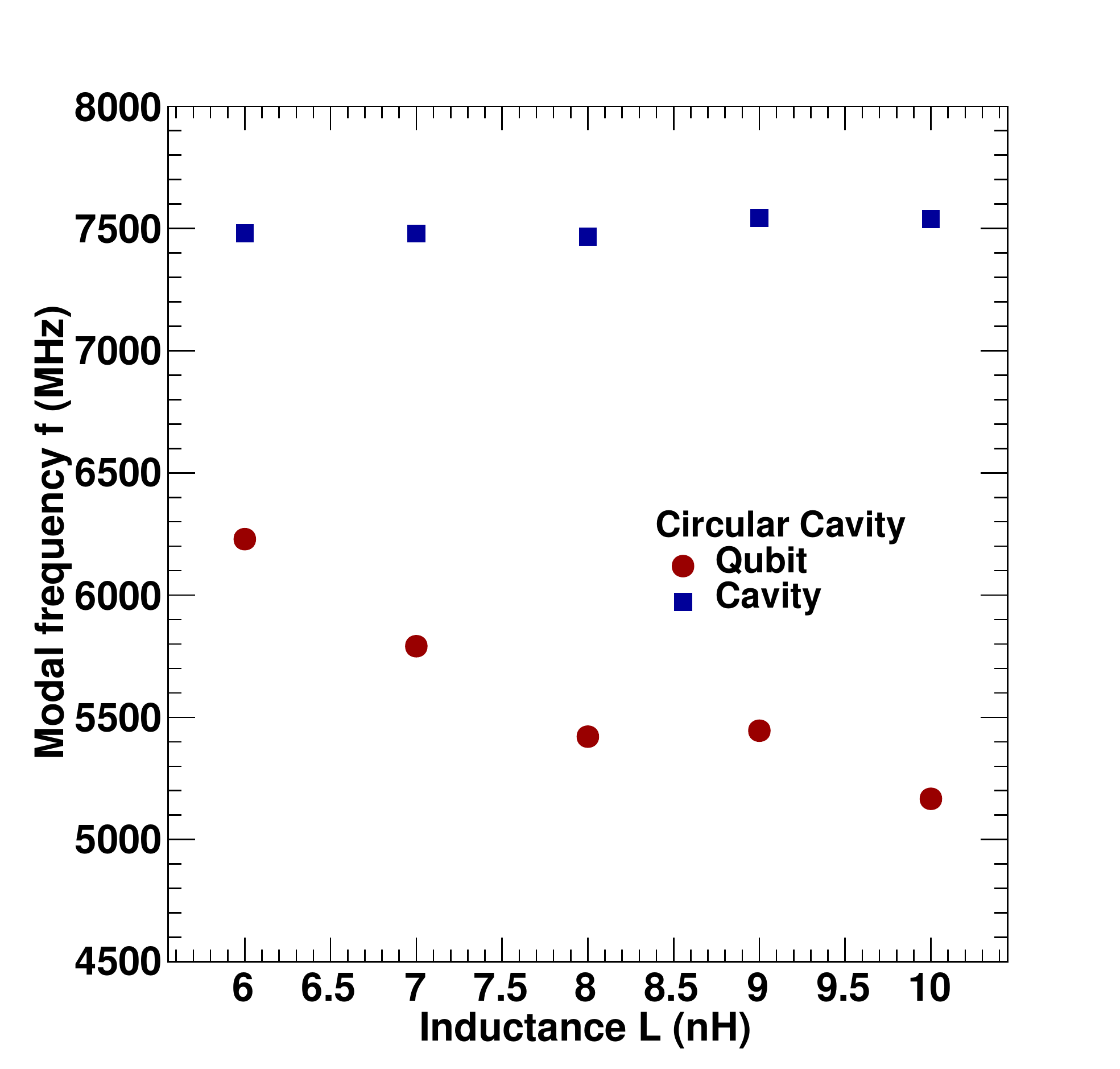}
  \includegraphics[width=0.45\textwidth]{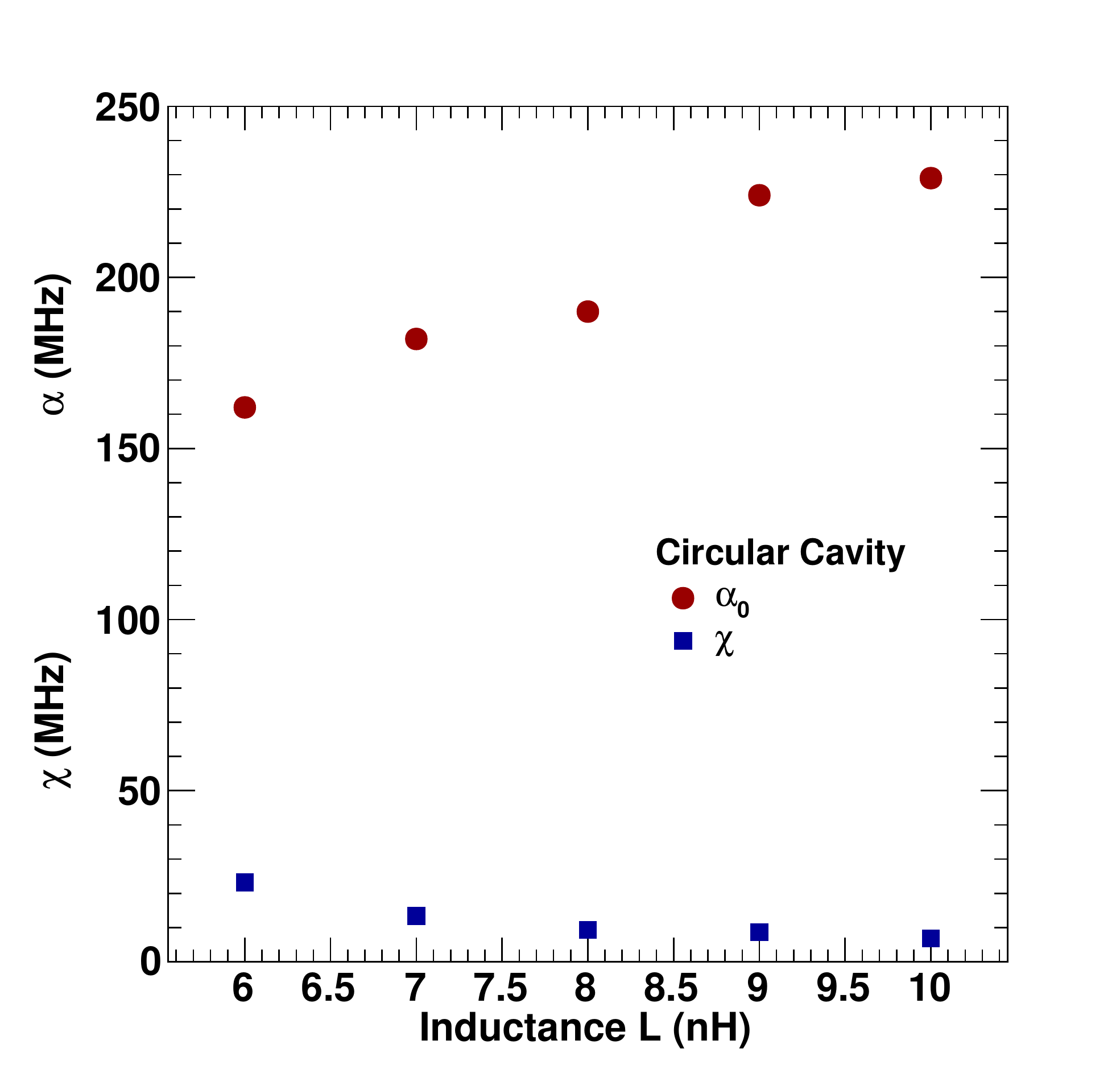}
  \caption{pyEPR simulations of a qubit coupled to a rectangular cavity showing modal 
frequencies (MHz), Anharmonicity $\alpha$ (MHz) and cross-kerr frequency $\chi$ (MHz) as a function of
different values of $L$.
}
\label{fig_cylcav1}
\end{figure}

Figure~\ref{fig_cylcav1} shows the
pyEPR simulations of a qubit coupled to a rectangular cavity giving modal 
frequencies (MHz), Anharmonicity $\alpha$ (MHz) and cross-kerr frequency $\chi$ (MHz) as a function of
different values of $L$.
For input parameter $L_J=6$ nH, $E_J=27.31$ GHz, 
the EPR value is 0.76 and 
$\omega_r/2\pi=7.481$ GHz, $\omega_q/2\pi=6.229$ GHz,
$\alpha/2\pi=162$ MHz, $\chi/2\pi=23.2$ MHz which gives $g/2\pi = \sqrt{\chi\triangle}=178$ MHz.

\begin{figure}
\centering
  \includegraphics[width=0.45\textwidth]{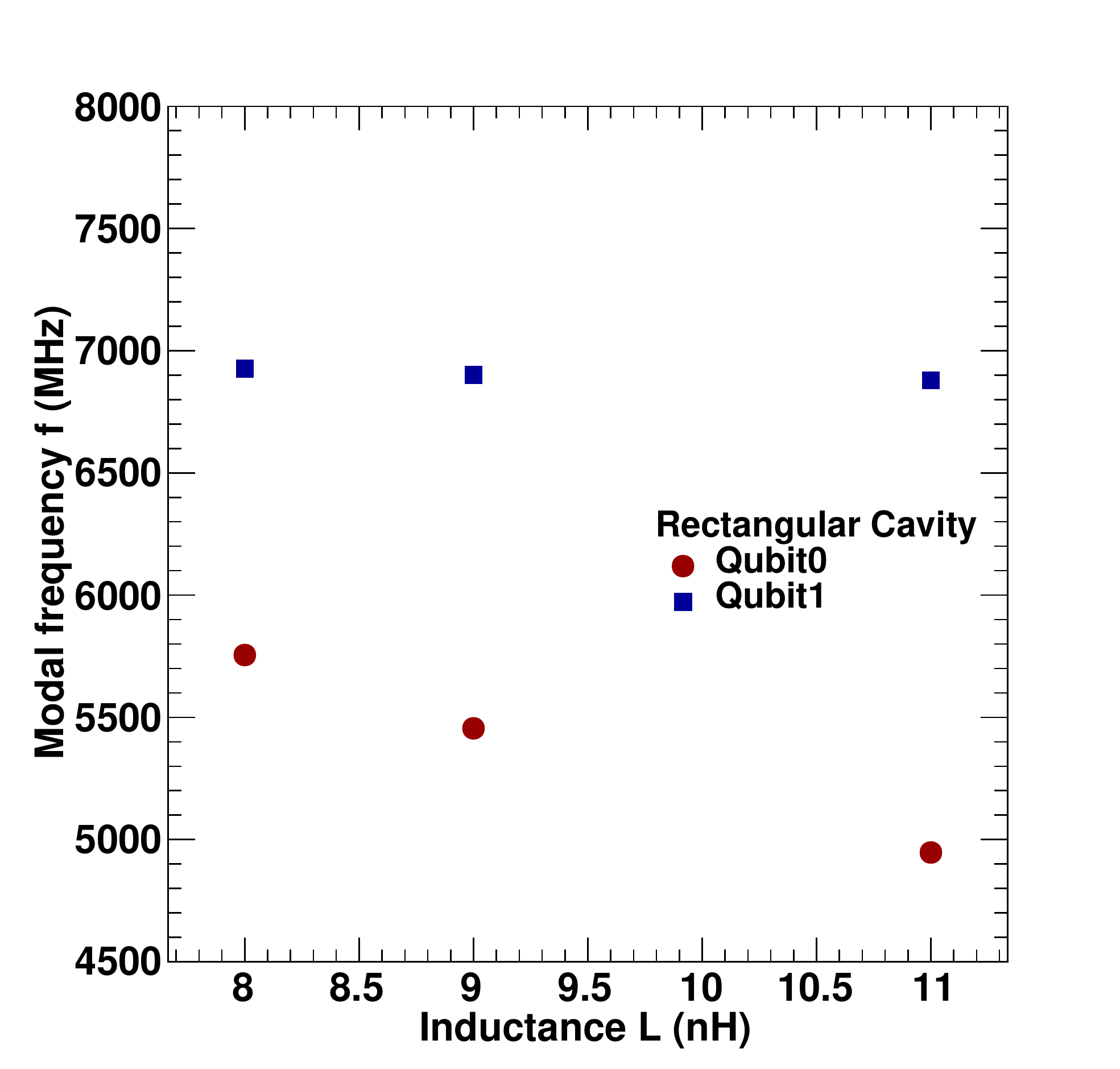}
  \includegraphics[width=0.45\textwidth]{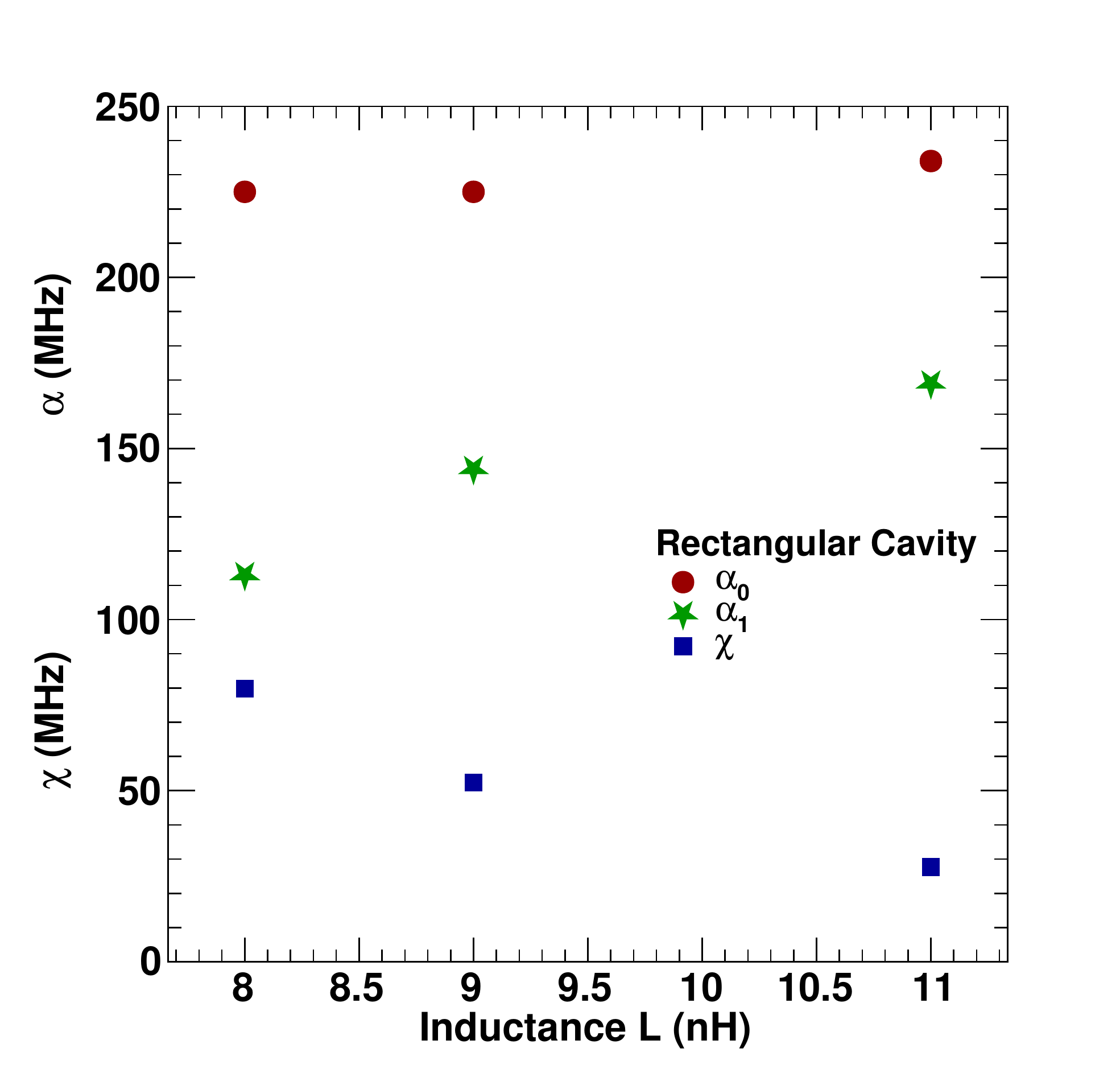}
\caption{pyEPR simulations of 2 coupled qubits showing modal frequencies (MHz), anharmonicities (MHz)
and cross-kerr frequencies (MHz). The variations correspond to the inductance of first junction
as 6 nH and different values of inductance of the second junction.}
\label{fig_qubit2}
\end{figure}

Figure~\ref{fig_qubit2} shows pyEPR simulations of 2 coupled qubits giving modal frequencies,
anharmonicities ($\alpha_0$, $\alpha_1$)  and cross-kerr frequency $\chi$.
The variations correspond to the inductance of first junction as 6 nH and different values of
inductance of the second junction.
Table~\ref{tab_qubit2} shows the results of pyEPR simulation of two junctions with
inductances 6 and 8 nH coupled to a rectangular cavity.


\begin{table}
\centering
\caption{The results of pyEPR simulation for a rectangular cavity coupled to
  two Josephson junctions for input inductance 6 and 8 nH.}
\begin{tabular}{ | c | c | c | c | c | }
\hline 
Mode   & $\omega_q/2\pi$ & $\alpha$  & $\chi_{01}/2\pi$   &  $g/2\pi = \sqrt{\chi_{01}\triangle}$ \\
\hline 
 0  &  6.926 GHz  &  225 MHz  &  79.8 MHz &  306  MHz \\
 1  &  5.755 GHz  &  113 MHz  &  79.8 MHz &  306  MHz \\
\hline
\end{tabular}
\label{tab_qubit2}
\end{table}

\section{Evolution of the open quantum system}
In the previous section, we got a quantitative idea about the Hamiltonian parameters like
frequencies, self-kerr, and cross-kerrs for our design.
Once Hamiltonian parameters are obtained for the circuit,
the energy levels of the system can be calculated. The dynamics of the system
can be obtained using Lindblad master equation \cite{Manzano_2020}. This yields frequency spectrum,
time spectrum, entanglement measures and calibration of microwave pulses.
The time evolution of the density matrix is given by Lindblad equation as follows
\begin{eqnarray}
  {\partial \over \partial t} \rho = -{i\over \hbar}[\Hcap(t), \rho(t)]  +
  \sum_{n} \half \left[2L_n\rho(t)L_n^+ - \rho(t) L_n^+L_n - L_n^+L_n \rho(t)\right].
\end{eqnarray}
Here, $\rho = \ket{\Psi}\bra{\Psi}$ is the density matrix and $L_n = \sqrt{\gamma_n} A_n$ are
collapse or jump operators.\\

Von Neumann entropy which is an extension of Gibbs entropy $S = k \ln Z$ is given by
\begin{eqnarray}
S  = -\sum_{i} \|  c_i \| ^2 \, \ln \|  c_i \| ^2 = - Tr(\rho\ln\rho).
\end{eqnarray}
If $ \ket{\Psi_{AB}} = \ket{\Psi_A}\ket{\Psi_B}$ is a separable state then entropy is zero.
Then the reduced density matrix
$\rho_A = Tr_B(\ket{\Psi_{AB}}\bra{\Psi_{AB}}) = \ket{\Psi_{A}}\bra{\Psi_{A}}$ is a pure state. 
The entanglement is characterized by non-zero entropy.


The qubit can relax in two ways; longitudinal and transverse relaxation. 
The longitudinal relaxation is actually energy decay and can be expressed 
as jump operators for resonator gaining or losing a photon
from a bath \cite{statistical2008}
\begin{eqnarray}
  L_{\uparrow} = \sqrt{\kappa_{\uparrow}}\, \ad, \,\,\,\,\,\,\,\,
  L_{\downarrow} = \sqrt{\kappa_{\downarrow}}\, \an . 
\end{eqnarray}
The ratio of excitation and relaxation rates can be obtained in terms of
Boltzmann factor 
\begin{eqnarray}
  \frac{\kappa_{\uparrow}}{\kappa_{\downarrow}} =   e^{-\hbar\omega/kT}.
\end{eqnarray}
Both the rates can be obtained in terms of single dissipation rate $\kappa_1$
and thermal occupancy $n_{th}$ as
\begin{eqnarray}
  \kappa_{\downarrow} = (1+n_{th}) \kappa_1, \,\,\,\,\,
  \kappa_{\uparrow} =  n_{th} \kappa_1 ,
\end{eqnarray}
where
\begin{eqnarray}
  n_{th} = \frac{1}{e^{\hbar\omega/kT} - 1}.
\end{eqnarray}
Similarly, for the qubit
\begin{eqnarray}
L_{q} = \sqrt{\Gamma_1}\, \sigma^-.
\end{eqnarray}

Decoherence or pure phase decay arises due to frequency change. 
The resonant frequency can be modified due to the scattering process of bath quantum
and the jump operator can be modeled as \cite{statistical2008}
\begin{eqnarray}
  L_{c\phi} = \sqrt{\kappa_\phi}\, \ad\an.
\end{eqnarray}
 For the qubit,
\begin{eqnarray}
  L_{q\phi} = \sqrt{\Gamma_\phi}\, \sigma^+\sigma^-.
\end{eqnarray}
The transverse relaxation includes longitudinal relaxation and pure dephasing
and are given for the resonator and the qubit as 
 \begin{eqnarray}
   \kappa_2 & = & \kappa_1/2 + \kappa_{\phi} \equiv \kappa , \nonumber \\
  \Gamma_2 & = & \Gamma_1/2 + \Gamma_{\phi} \equiv \Gamma.
\end{eqnarray}

\begin{figure} 
\centering
  \includegraphics[width=0.65\textwidth]{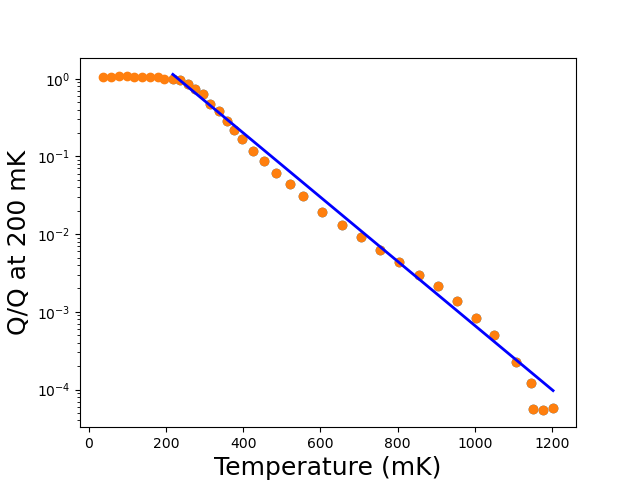}
 \caption{The ratio of quality factor of aluminium cavity with the quality factor at 200 mK
   as a function of temperature~\cite{Reagor_2013} along
  with an exponential function.}
\label{fig_QFcav}
\end{figure}

 The dissipation in the cavity varies as a function of temperature.
Figure~\ref{fig_QFcav} shows the ratio of quality factor of aluminum cavity with
the quality factor at 200 mK as a function of temperature~\cite{Reagor_2013}.
The quality factor has been fitted by an exponential function
\begin{eqnarray}
  Q/Q_{200mK} & = & 9.06 \, \exp(-T/105) \,\,\, (T\, {\rm in\, mK}) .
\end{eqnarray}
From this, we can get the quality factor at 1 K as follows
\begin{eqnarray}
  Q_{1000mK}  & = & 0.00015 \times Q_{200mK} \nonumber \\ 
            & = & 0.00015 \times (7 \times 10^7) = 10500 .
\end{eqnarray}
This corresponds to longitudinal dissipation in cavity as 
$\kappa_1/(2\pi) = \nu/Q = 0.5$ MHz. 
Assuming pure dephasing rate as 0.25 MHz the transverse relaxation 
$\kappa_2/(2\pi) = \nu/(2Q) = 0.25$ MHz + 0.25 MHz. The dissipations in superconducting
qubits are of the order of $\Gamma/2\pi$ = $0.01$ MHz \cite{Reagor_2016}.

\begin{figure}
\centering
  \includegraphics[width=0.55\textwidth]{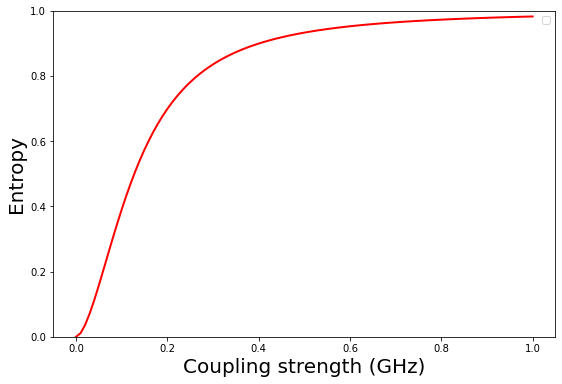}
 \caption{Entropy for two qubits entanglement as a function of coupling strength
 for $\kappa/2\pi=10^{-5}$ MHz (cavity), $\Gamma/2\pi=10^{-5}$ MHz and 
   Temperature at 200 mK. }
\label{fig_entropy12}
\end{figure}

Figure~\ref{fig_entropy12} shows entropy for two qubits entanglement as a function
of coupling strength for $\kappa/2\pi=10^{-5}$ MHz (cavity), $\Gamma/2\pi=10^{-5}$ MHz and 
Temperature at 200 mK. Table~\ref{tab_qubit2} shows that the coupling between the two
qubits obtained is 306 MHz which corresponds to entropy 0.85 in Fig.~\ref{fig_entropy12}
showing a high degree of entanglement.

\section{Rabi Oscillation}
\begin{figure}
\centering
  \includegraphics[width=0.70\textwidth]{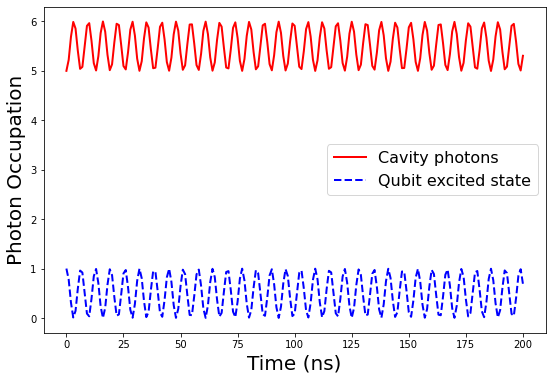}
\caption{Rabi oscillations for a system for parameters
$\omega_r/2\pi$ = 7 GHz,  $\omega_q/2\pi$ = 7 GHz,  coupling $g$ = 200 MHz,
$\kappa/2\pi$ = $10^{-5}$ MHz (cavity), $\Gamma/2\pi$ = $0.01$ MHz and $T$ = 200 mK.}
\label{fig_Rb1_200}
\end{figure}

In this section, we present the results of the Rabi oscillation \cite{Wallraff2004}
in a qubit-cavity system calculated using Lindablad equations with a set of realistic parameters. 
Rabi oscillations for a qubit-cavity system for parameters
$\omega_r/2\pi$ = 7 GHz,  $\omega_q/2\pi$ = 7 GHz,  coupling $g$ = 200 MHz,
$\kappa/2\pi$ = $10^{-5}$ MHz (cavity), $\Gamma/2\pi$ = $0.01$ MHz and $T$ = 200 mK
are given in Fig.~\ref{fig_Rb1_200}. Here we start with 5 photons in the cavity and one photon
is exchanged between the qubit and cavity. It shows that the Rabi oscillations can
be very well observed at 200 mK. 

\begin{figure}
\centering
  \includegraphics[width=0.70\textwidth]{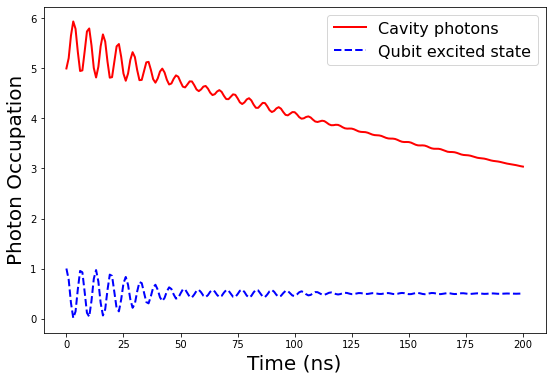}
\caption{Rabi oscillations for a system for parameters
$\omega_r/2\pi$ = 7 GHz,  $\omega_q/2\pi$ = 7 GHz,  coupling $g$ = 200 MHz,
$\kappa/2\pi$ = 0.5 MHz (cavity) and $\Gamma/2\pi$ = $0.01$ MHz and $T$ = 200 mK.}
\label{fig_Rb2_200}
\end{figure}

Rabi oscillations for a qubit-cavity system for parameters
$\omega_r/2\pi$ = 7 GHz,  $\omega_q/2\pi$ = 7 GHz,  coupling $g$ = 200 MHz,
$\kappa/2\pi$ = 0.5 MHz (cavity), $\Gamma/2\pi$ = $0.01$ MHz and $T$ = 200 mK
are given in Fig.~\ref{fig_Rb2_200}. Here we have increased the dissipation in the cavity
corresponding to $\kappa/2\pi$ = 0.5 MHz which has made the oscillations die down
within 100 ns. 

\begin{figure}
\centering
  \includegraphics[width=0.70\textwidth]{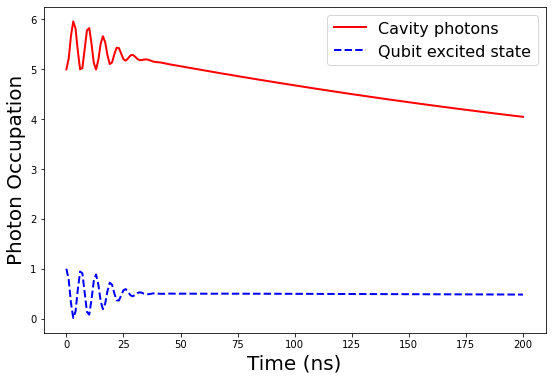}
\caption{Rabi oscillations for a system for parameters
$\omega_r/2\pi$ = 7 GHz,  $\omega_q/2\pi$ = 7 GHz,  coupling $g$ = 200 MHz,
$\kappa/2\pi$ = 0.5 MHz (cavity), $\Gamma/2\pi$ = $0.01$ MHz and $T$ = 1 K.}
\label{fig_Rb3_200}
\end{figure}

Rabi oscillations for a system for parameters
$\omega_r/2\pi$ = 7 GHz,  $\omega_q/2\pi$ = 7 GHz,  coupling $g$ = 200 MHz,
$\kappa/2\pi$ = 0.5 MHz (cavity), $\Gamma/2\pi$ = $0.01$ MHz and $T$ = 1 K
are given in Fig.~\ref{fig_Rb3_200}.
 Here we have added dissipation in the cavity
 corresponding to $\kappa/2\pi$ = 0.5 MHz and also increased the temperature to
 1 K  which has made the oscillations die down within 30 ns. 

\section{Driven qubit}
\begin{figure}
\centering  
 \includegraphics[width=0.70\textwidth]{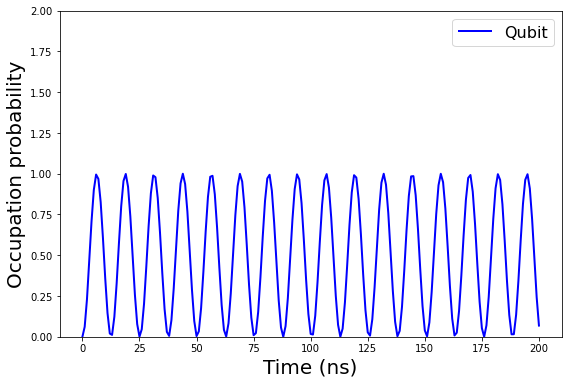}
\caption{Rabi oscillations driven by a microwave signal with coupling strength
  $g \sim 500$ MHz for $\omega_q=\omega_d$.}
\label{fig_Rb0}
\end{figure}

Hamiltonian for a qubit driven by a microwave signal is \cite{Naghiloo_2019}
\begin{equation}
H = -{1\over 2} \omega_q \, \sigma_z + \Omega\, V_d(t) \sigma_y.
\end{equation}
Applying $V_d(t) = V_0 \cos(\omega_dt)$ and defining $g=\Omega V_0$ MHz we obtain
\begin{eqnarray*}
  \Hcap = - \half w_q\sigma_z + g\,\cos(\omega_d t)\sigma_y.
\end{eqnarray*}
The probability of qubit being in an excited state is obtained as 
\begin{eqnarray*}
  P_e(t) = {g^2 \over \Omega_R^2} \,   \sin^2\left(\Omega_Rt \over 2\right).
\end{eqnarray*}
Here, $\Omega_R = \sqrt{g^2+(\omega_q-\omega_d)^2}$.
Figure~\ref{fig_Rb0} shows the Rabi oscillations driven by a microwave signal
with coupling strength $g \sim 500$ MHz for $\omega_q=\omega_d$ .

\begin{figure}
\centering
  \includegraphics[width=0.40\textwidth]{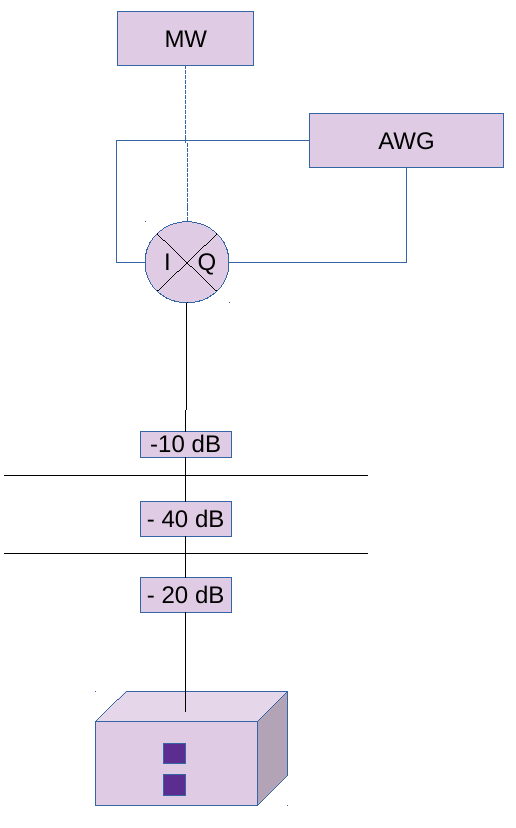} 
\caption{A setup with microwave drive mixed with an envelope function and given to qubit.}
\label{fig_setup1}
\end{figure}

\begin{figure}
\centering
  \includegraphics[width=0.60\textwidth]{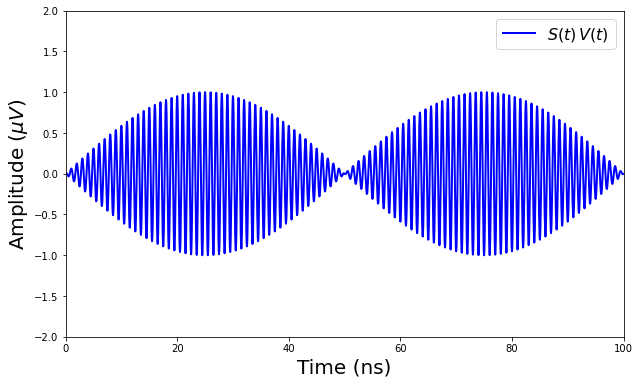}
\caption{The microwave pulse with a sine envelope function with frequency 10 MHz.}
\label{fig_pulse1}
\end{figure}

Figure~\ref{fig_setup1} shows a setup with a microwave (MW) drive mixed with
an envelope function $S(t)$ generated using Arbitrary Wave Generator (AWG)
and given to qubit. The resultant drive voltage is
\begin{equation}
  V_d(t) = S(t) \, V_0\, \sin (\omega_d t +  \phi).
\end{equation}
Figure~\ref{fig_pulse1} shows a microwave pulse ($f=\omega_d/2\pi$=1 GHz for demonstration) with a
sine envelope function with a frequency 10 MHz.
The envelope function $S(t)=\sin(\omega_et) $ with $\omega_e=10$ MHz is used in the calculations.

\begin{figure}
\centering
  \includegraphics[width=0.48\textwidth]{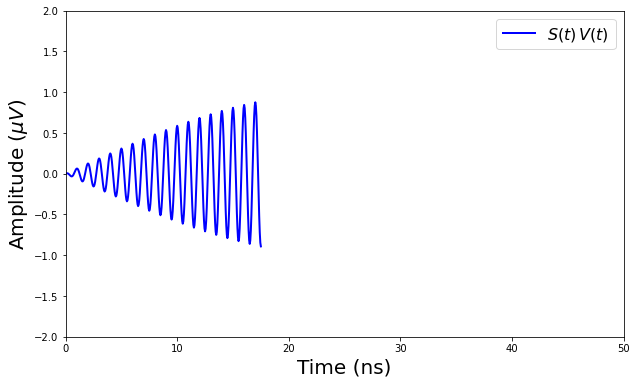}
  \includegraphics[width=0.48\textwidth]{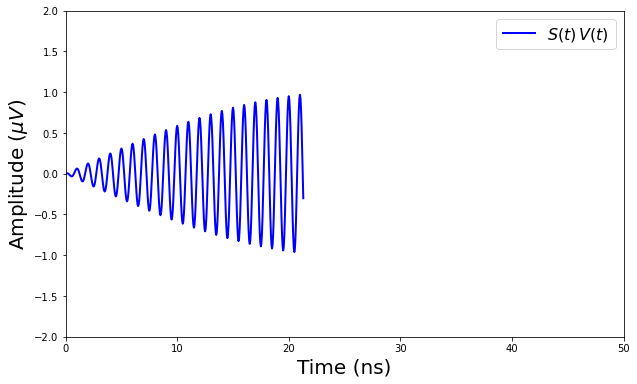}
  \caption{The $\pi$ pulse (duration $t = 17.6$ ns) which can be used for reversing
    the state of the qubit and $\pi/2$ pulse ($t =21.4$ ns) is used to put it in a superposition state. 
}
\label{fig_pulsepi}
\end{figure}

The qubit can be manipulated using microwave pulses.
The time duration of pulse for qubit rotation $\Theta$ can be obtained by solving 
 
\begin{equation}
  \Theta(t) = -g \int_0^t S(t') dt'.
\end{equation}
Figure~\ref{fig_pulsepi} shows the $\pi$ pulse (duration $t = 17.6$ ns) which
can be used for reversing
the state of the qubit and $\pi/2$ pulse ($t =21.4$ ns) is used to put it in a superposition state.

\section{Measurement of frequency shift}

\begin{figure}
\centering
  \includegraphics[width=0.40\textwidth]{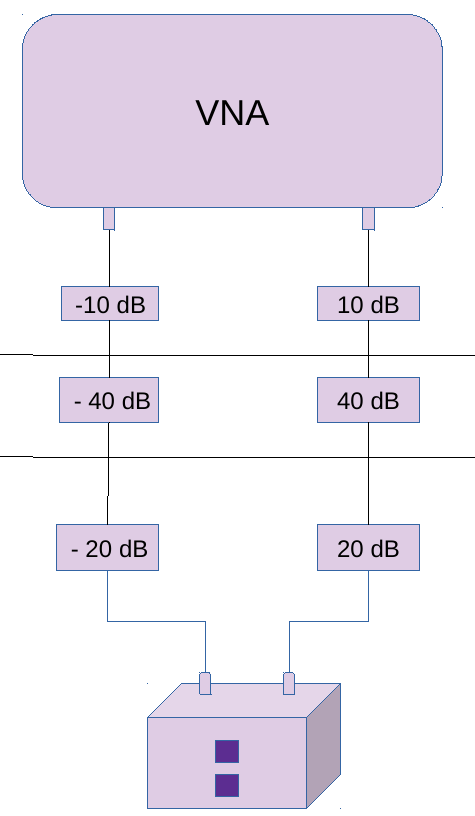}
\caption{An experimental setup for frequency sweep.}
\label{fig_setup2}
\end{figure}

Figure~\ref{fig_setup2} shows an experimental setup for frequency sweep.
Here a signal from Vector Network Analyser (VNA) is
sent to qubit and frequency sweep is done. The transmitted signal (S21) is
measured via the 2nd port. 

\begin{figure}
\centering
  \includegraphics[width=0.49\textwidth]{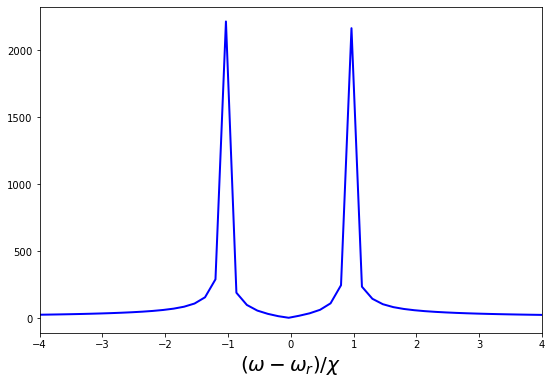}
  \includegraphics[width=0.49\textwidth]{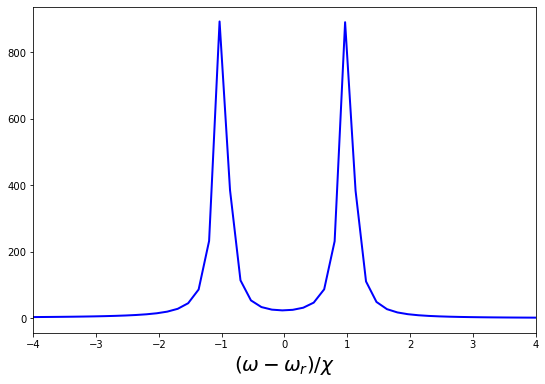}
\caption{Frequency shift of the cavity due to qubit for
  $\triangle=$ 1 GHz, $\chi/2\pi=$ 6 MHz and temperature = 200 mK.
  Left figure is for $\kappa/2\pi=$ 0.1 MHz and the right figure is
  for $\kappa/2\pi=$ 0.5 MHz.
}   
\label{fig_chi01}
\end{figure}

Figure~\ref{fig_chi01} shows the 
frequency shift of the cavity due to qubit for
  $\triangle=$ 1 GHz, $\chi/2\pi=$ 6 MHz and temperature = 200 mK.
  Left figure is for $\kappa/2\pi=$ 0.1 MHz and the right figure is
  for $\kappa/2\pi=$ 0.5 MHz. It is observed that the width of the lines
  is increased but they are well separated and will give the measure of
  qubit frequency.

\begin{figure}
\centering
  \includegraphics[width=0.49\textwidth]{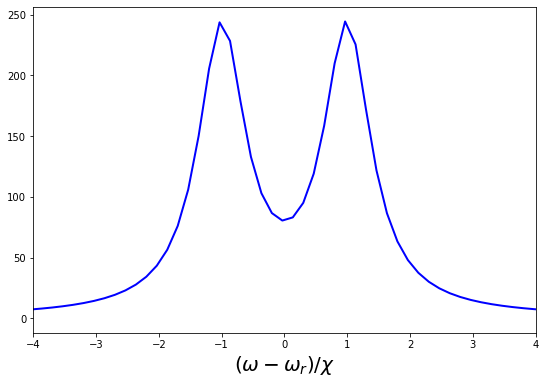}
  \includegraphics[width=0.49\textwidth]{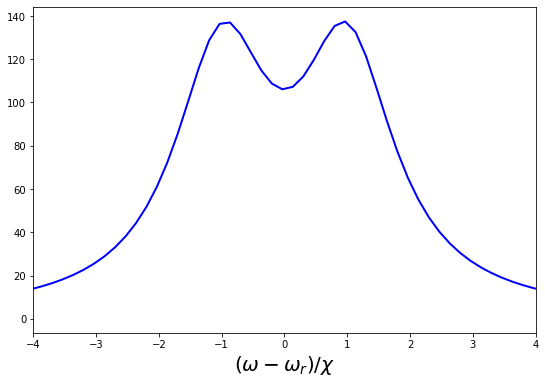}
  \caption{Frequency shift of the cavity due to qubit
for $\triangle=$ 1 GHz, $\chi/2\pi=$ 6 MHz and temperature = 1000 mK.
Left figure is for $\kappa/2\pi=$ 0.5 MHz and right figure is for
$\kappa/2\pi=$ 1 MHz.
}
\label{fig_chi02}
\end{figure}

Figure~\ref{fig_chi02} shows the frequency shift of the cavity due to qubit
for $\triangle=$ 1 GHz, $\chi/2\pi=$ 6 MHz and temperature = 1000 mK.
Left figure is for $\kappa/2\pi=$ 0.5 MHz and right figure is for $\kappa/2\pi=$ 1 MHz.
 It is observed that the width of the lines
 is increased but the difference can still be measured. For dissipation
 beyond $\kappa/2\pi>$ 1 MHz, the two lines merge. We can conclude that it is
 still possible to measure the frequency shift if the quality of the resonator is good.  
  
\section{Summary}

 We presented a study of the dynamics of superconducting qubits coupled with
cavity by including dissipation in sub kelvin temperature range.  
All the steps starting with a classical simulation, EPR method and
Lindblad equation have been described. Once we have the parameters of Hamiltonian
we can solve a quantum mechanical equation. 
An open quantum system formalism is implemented to study the effect
of dissipation and finite temperature on the dynamics. 
Calculations are performed using a realistic set of dissipative parameters
and include temperatures up to 1 K. 
Finally, we get frequency spectra and/or dynamics of the system with time
with quantum effects.
Such devices work at tens of milli Kelvins and quantum effects diminish as we
move to higher temperatures. 
 We find a set of parameters for which one can observe quantum behaviour
up to 1 K for a resonator with a fairly good quality factor.

\section*{Data Availability Statement}

The datasets generated during and/or analysed during the current study are
available from the corresponding author on reasonable request.

\bibliography{DissPaper.bib}


\end{document}